\RequirePackage{lineno}
\RequirePackage{xspace}

\documentclass[aps,prl,twocolumn,superscriptaddress,showpacs,preprintnumbers,amsmath,amssymb]{revtex4-2}

\usepackage{graphicx} 
\usepackage{dcolumn}  
\usepackage{makecell}

\allowdisplaybreaks
\graphicspath{{ps}}

\input belle2sym.tex

\begin{document}

\title{ \quad\\[1.0cm] \boldmath Search for \Bz meson decays into $\Lambda$ and missing energy\\with a hadronic tagging method at Belle}

\noaffiliation
\affiliation{Department of Physics, University of the Basque Country UPV/EHU, 48080 Bilbao}
\affiliation{University of Bonn, 53115 Bonn}
\affiliation{Brookhaven National Laboratory, Upton, New York 11973}
\affiliation{Budker Institute of Nuclear Physics SB RAS, Novosibirsk 630090}
\affiliation{Faculty of Mathematics and Physics, Charles University, 121 16 Prague}
\affiliation{University of Cincinnati, Cincinnati, Ohio 45221}
\affiliation{Deutsches Elektronen--Synchrotron, 22607 Hamburg}
\affiliation{University of Florida, Gainesville, Florida 32611}
\affiliation{Department of Physics, Fu Jen Catholic University, Taipei 24205}
\affiliation{Key Laboratory of Nuclear Physics and Ion-beam Application (MOE) and Institute of Modern Physics, Fudan University, Shanghai 200443}
\affiliation{Gifu University, Gifu 501-1193}
\affiliation{SOKENDAI (The Graduate University for Advanced Studies), Hayama 240-0193}
\affiliation{Gyeongsang National University, Jinju 52828}
\affiliation{Department of Physics and Institute of Natural Sciences, Hanyang University, Seoul 04763}
\affiliation{University of Hawaii, Honolulu, Hawaii 96822}
\affiliation{High Energy Accelerator Research Organization (KEK), Tsukuba 305-0801}
\affiliation{J-PARC Branch, KEK Theory Center, High Energy Accelerator Research Organization (KEK), Tsukuba 305-0801}
\affiliation{National Research University Higher School of Economics, Moscow 101000}
\affiliation{Forschungszentrum J\"{u}lich, 52425 J\"{u}lich}
\affiliation{IKERBASQUE, Basque Foundation for Science, 48013 Bilbao}
\affiliation{Indian Institute of Science Education and Research Mohali, SAS Nagar, 140306}
\affiliation{Indian Institute of Technology Guwahati, Assam 781039}
\affiliation{Indian Institute of Technology Hyderabad, Telangana 502285}
\affiliation{Indian Institute of Technology Madras, Chennai 600036}
\affiliation{Indiana University, Bloomington, Indiana 47408}
\affiliation{Institute of High Energy Physics, Chinese Academy of Sciences, Beijing 100049}
\affiliation{Institute of High Energy Physics, Vienna 1050}
\affiliation{Institute for High Energy Physics, Protvino 142281}
\affiliation{INFN - Sezione di Napoli, I-80126 Napoli}
\affiliation{INFN - Sezione di Roma Tre, I-00146 Roma}
\affiliation{INFN - Sezione di Torino, I-10125 Torino}
\affiliation{Advanced Science Research Center, Japan Atomic Energy Agency, Naka 319-1195}
\affiliation{J. Stefan Institute, 1000 Ljubljana}
\affiliation{Institut f\"ur Experimentelle Teilchenphysik, Karlsruher Institut f\"ur Technologie, 76131 Karlsruhe}
\affiliation{Kitasato University, Sagamihara 252-0373}
\affiliation{Korea University, Seoul 02841}
\affiliation{Kyungpook National University, Daegu 41566}
\affiliation{Universit\'{e} Paris-Saclay, CNRS/IN2P3, IJCLab, 91405 Orsay}
\affiliation{P.N. Lebedev Physical Institute of the Russian Academy of Sciences, Moscow 119991}
\affiliation{Faculty of Mathematics and Physics, University of Ljubljana, 1000 Ljubljana}
\affiliation{Ludwig Maximilians University, 80539 Munich}
\affiliation{Luther College, Decorah, Iowa 52101}
\affiliation{Malaviya National Institute of Technology Jaipur, Jaipur 302017}
\affiliation{Faculty of Chemistry and Chemical Engineering, University of Maribor, 2000 Maribor}
\affiliation{Max-Planck-Institut f\"ur Physik, 80805 M\"unchen}
\affiliation{School of Physics, University of Melbourne, Victoria 3010}
\affiliation{University of Mississippi, University, Mississippi 38677}
\affiliation{Moscow Physical Engineering Institute, Moscow 115409}
\affiliation{Graduate School of Science, Nagoya University, Nagoya 464-8602}
\affiliation{Kobayashi-Maskawa Institute, Nagoya University, Nagoya 464-8602}
\affiliation{Universit\`{a} di Napoli Federico II, I-80126 Napoli}
\affiliation{Nara Women's University, Nara 630-8506}
\affiliation{National Central University, Chung-li 32054}
\affiliation{National United University, Miao Li 36003}
\affiliation{Department of Physics, National Taiwan University, Taipei 10617}
\affiliation{H. Niewodniczanski Institute of Nuclear Physics, Krakow 31-342}
\affiliation{Nippon Dental University, Niigata 951-8580}
\affiliation{Niigata University, Niigata 950-2181}
\affiliation{Novosibirsk State University, Novosibirsk 630090}
\affiliation{Osaka City University, Osaka 558-8585}
\affiliation{Pacific Northwest National Laboratory, Richland, Washington 99352}
\affiliation{Panjab University, Chandigarh 160014}
\affiliation{Peking University, Beijing 100871}
\affiliation{University of Pittsburgh, Pittsburgh, Pennsylvania 15260}
\affiliation{Punjab Agricultural University, Ludhiana 141004}
\affiliation{Research Center for Nuclear Physics, Osaka University, Osaka 567-0047}
\affiliation{Meson Science Laboratory, Cluster for Pioneering Research, RIKEN, Saitama 351-0198}
\affiliation{Department of Modern Physics and State Key Laboratory of Particle Detection and Electronics, University of Science and Technology of China, Hefei 230026}
\affiliation{Showa Pharmaceutical University, Tokyo 194-8543}
\affiliation{Soongsil University, Seoul 06978}
\affiliation{Sungkyunkwan University, Suwon 16419}
\affiliation{School of Physics, University of Sydney, New South Wales 2006}
\affiliation{Tata Institute of Fundamental Research, Mumbai 400005}
\affiliation{Department of Physics, Technische Universit\"at M\"unchen, 85748 Garching}
\affiliation{School of Physics and Astronomy, Tel Aviv University, Tel Aviv 69978}
\affiliation{Department of Physics, Tohoku University, Sendai 980-8578}
\affiliation{Earthquake Research Institute, University of Tokyo, Tokyo 113-0032}
\affiliation{Department of Physics, University of Tokyo, Tokyo 113-0033}
\affiliation{Tokyo Institute of Technology, Tokyo 152-8550}
\affiliation{Tokyo Metropolitan University, Tokyo 192-0397}
\affiliation{Virginia Polytechnic Institute and State University, Blacksburg, Virginia 24061}
\affiliation{Wayne State University, Detroit, Michigan 48202}
\affiliation{Yamagata University, Yamagata 990-8560}
\affiliation{Yonsei University, Seoul 03722}
  \author{C.~Hadjivasiliou}\affiliation{Pacific Northwest National Laboratory, Richland, Washington 99352} 
  \author{B.~G.~Fulsom}\affiliation{Pacific Northwest National Laboratory, Richland, Washington 99352} 
  \author{J.~F.~Strube}\affiliation{Pacific Northwest National Laboratory, Richland, Washington 99352} 
  \author{I.~Adachi}\affiliation{High Energy Accelerator Research Organization (KEK), Tsukuba 305-0801}\affiliation{SOKENDAI (The Graduate University for Advanced Studies), Hayama 240-0193} 
  \author{H.~Aihara}\affiliation{Department of Physics, University of Tokyo, Tokyo 113-0033} 
  \author{D.~M.~Asner}\affiliation{Brookhaven National Laboratory, Upton, New York 11973} 
  \author{H.~Atmacan}\affiliation{University of Cincinnati, Cincinnati, Ohio 45221} 
  \author{T.~Aushev}\affiliation{National Research University Higher School of Economics, Moscow 101000} 
  \author{V.~Babu}\affiliation{Deutsches Elektronen--Synchrotron, 22607 Hamburg} 
  \author{K.~Belous}\affiliation{Institute for High Energy Physics, Protvino 142281} 
% ...
\collaboration{The Belle Collaboration}

\begin{abstract}
We present a search for the decays of \Bz mesons into a final state containing a $\Lambda$ baryon and missing energy. These results are obtained from a $711\invfb$ data sample  that contains $772 \times 10^6$ \BB pairs and was collected near the \FourS resonance with the Belle detector at the KEKB asymmetric-energy \epem collider. We use events in which one \B meson is fully reconstructed in a hadronic decay mode and require the remainder of the event to consist of only a single $\Lambda$. No evidence for these decays is found and we set $90\%$ confidence level upper limits on the branching fractions in the range $2.1$--$3.8\times 10^{-5}$. This measurement provides the world's most restrictive limits, with implications for baryogenesis and dark matter production.
\end{abstract}

\maketitle


\tighten

{\renewcommand{\thefootnote}{\fnsymbol{footnote}}}
\setcounter{footnote}{0}

\begin{figure}[htb]
\includegraphics[width=0.45\textwidth]{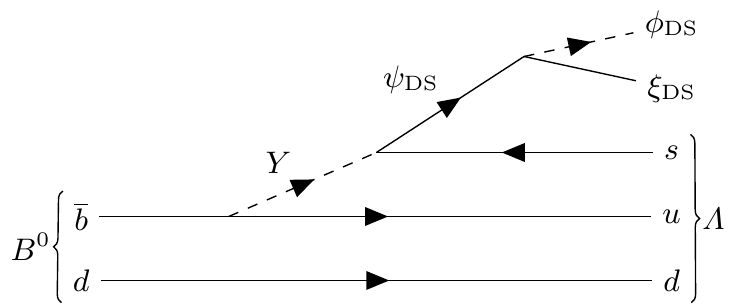}
\caption{An example diagram of the \B meson decay process as mediated by the heavy colored scalar $Y$ that results in DM and a visible baryon. The dark sector antibaryon $\psiDM$ decays into stable DM particles: a dark sector scalar antibaryon $\phi_\text{DS}$ and a dark Majorana fermion $\xi_\text{DS}$.}
\label{fig:bdecay}
\end{figure}

According to the \B-Mesogenesis mechanism~\cite{Elor:2018twp, Alonso-Alvarez:2021qfd}, the {\it CP}-violating oscillations and subsequent decays of \B mesons in the early Universe can simultaneously explain the dark matter (DM) relic abundance and baryon asymmetry. A robust prediction of this mechanism is a branching fraction larger than ${\cal B}_\text{M}=10^{-4}$ for \Bz mesons decaying into a final state containing a $\Lambda$ baryon, missing energy in the form of a \gev-scale dark sector antibaryon $\psiDM$, and any number of light mesons; $\BFDMm>10^{-4}$. The limit ${\cal B}_\text{M}$ strongly depends on the semileptonic asymmetries in neutral \B meson decays~\cite{Alonso-Alvarez:2021qfd, Zyla:2020zbs}. At present, the best bound on such a process is an exclusive branching fraction of $\BFDM\lesssim 2\times 10^{-4}$ derived from an inclusive ALEPH search for events with large missing energy arising from $b$-flavored hadron decays at the $Z$ peak~\cite{ALEPH, Alonso-Alvarez:2021qfd}. In order for this decay to exist, a new \tev-scale bosonic colored mediator $Y$ is required. This heavy mediator can be integrated out to yield an effective four-fermion operator ${\cal O}_{us}=\psiDM bus$. An example diagram of the corresponding decay is shown in Fig.~\ref{fig:bdecay}. Successful baryogenesis requires a $\psiDM$ mass $\lesssim 3.5\gev/c^2$ as indirectly constrained by LHC searches on \tev-scale color-triplet scalars~\cite{LHC:squark, Alonso-Alvarez:2021qfd}. We report the first search for \BToLDM exclusive decays using the full Belle data sample of $711\invfb$ collected near the \FourS resonance. Charge-conjugate decays are implied throughout this letter.

This measurement is based on a data sample that contains $(772\pm11)\times 10^6$ \BB pairs, collected  with the Belle detector~\cite{Belle} at the KEKB asymmetric-energy \epem ($3.5$ on $8.0\gev$) collider~\cite{KEKB} operated at the \FourS resonance. In addition, we employ an $89\invfb$ data sample recorded at a center-of-mass (CM) energy $60\mev$ below the \FourS resonance (off-resonance data) to characterize the background. The Belle detector is a large-solid-angle magnetic spectrometer that consists of a silicon vertex detector (SVD), a 50-layer central drift chamber (CDC), an array of aerogel threshold Cherenkov counters (ACC), a barrel-like arrangement of time-of-flight scintillation counters (TOF), and an electromagnetic calorimeter comprised of CsI(Tl) crystals (ECL) located inside a superconducting solenoid coil that provides a $1.5\ensuremath{{\rm \,T}}\xspace$ magnetic field. An iron flux-return located outside of the coil is instrumented to detect \KL mesons and to identify muons. Two inner detector configurations were used. A $2.0\cm$ radius beampipe and a 3-layer SVD were used for the first sample of $152 \times 10^6$ \BB pairs, while a $1.5\cm$ radius beampipe, a 4-layer SVD, and a small-inner-cell CDC were used to record the remaining $620 \times 10^6$ \BB pairs~\cite{Natkaniec:2006rv}.

We study properties of signal events, identify sources of background, and optimize selection criteria using Monte Carlo (MC) simulated events. These samples are generated using the software packages \textsc{EvtGen}~\cite{Lange:2001uf} and \textsc{Pythia}~\cite{Sjostrand:2006za}, and final-state radiation is included via \textsc{Photos}~\cite{Barberio:1990ms}. The detector response is simulated using \textsc{Geant3}~\cite{Brun:1987ma}. We produce \BToLDM MC events according to a phase-space model for eight individual values of the $\psiDM$ mass in the range $1.0\gev/c^2\le m_{\psiDM}\le 3.9\gev/c^2$ to calculate signal reconstruction efficiencies. To estimate background, we use MC samples that describe all $\epem\to\qqbar$ processes. Events containing $\epem\to\BB$ with subsequent $b\to c$ decays, and $\epem\to\qqbar$ ($q=u,d,s,c$) continuum events, are both simulated with six times the integrated luminosity of Belle. Rare charmless \B meson decays are simulated with 50 times the integrated luminosity. 

Event reconstruction for this analysis is performed entirely in the Belle II Software Framework~\cite{Kuhr:2018lps} by converting Belle data structures to that of Belle II~\cite{Gelb:2018agf}. We identify signal candidates by fully reconstructing the accompanying neutral \B meson ($B_\text{tag}$) and requiring a single $\Lambda$ baryon on the signal side. We analyze the data in an unbiased manner by finalizing all selection criteria before viewing events in the signal region.

The $B_\text{tag}$ candidates are reconstructed in hadronic decay channels using the Full Event Interpretation algorithm~\cite{Keck:2018lcd}. The algorithm employs a hierarchical reconstruction ansatz in six stages. In the first stage, tracks and neutral clusters are identified and required to pass some basic quality criteria. In the second stage, boosted decision trees (BDTs) are trained to identify charged tracks and neutral energy depositions as detector-stable particles ($\pip, \Kp, \mup, \ep, \gamma$). In the third and fourth stages, these candidate particles are combined into composite parents ($\piz, \jpsi, \KS, \Dz, \Dp, \Ds$), and for each target final state, a BDT is trained to identify probable candidates. At the fifth stage, candidates for excited mesons ($\Dstarz, \Dstarp, \Dss$) are formed and separate BDTs are trained to identify viable combinations. The input variables of each stage aggregate the output classifiers from all previous reconstruction stages. The final stage combines the information from all previous stages to form $B_\text{tag}$ candidates. The viability of such combinations is assessed by a BDT that is trained to distinguish correctly reconstructed candidates from wrong combinations, and whose output classifier score we denote as the signal probability $o_\text{tag}$. The purity of the $B_\text{tag}$ candidate is improved by selecting candidates with $o_\text{tag}>10^{-3}$. We further select the $B_\text{tag}$ candidates using the energy difference $\Delta E\equiv E_B-E_\text{beam}$ and the beam-energy constrained mass $M_\text{bc}\equiv\sqrt{E_\text{beam}^2/c^4-{\lvert{{\vec{p}_B}}\rvert^2}/c^2}$, where $E_B$ and $\vec{p}_B$ are the reconstructed energy and momentum of the $B_\text{tag}$ candidate in the $\FourS$ CM frame, and $E_\text{beam}$ is the beam energy in this frame. We require $B_\text{tag}$ candidates to satisfy the requirements $5.27\gev/c^2<M_\text{bc}<5.29\gev/c^2$ and $\left|\Delta E\right|<0.06\gev$. We apply a calibration factor for the hadronic tagging efficiency, derived from exclusive measurements of $\Bz\to Xl\nu$ decay channels, to all correctly reconstructed $B_\text{tag}$ candidates in MC simulation. A full description of this procedure can be found in Ref.~\cite{Keck:2018lcd}.

The particles in the event not associated with the $B_\text{tag}$ meson are used to reconstruct a $B_\text{sig}\to\Lambda\psiDM$ candidate. The $\Lambda$ candidates are reconstructed via $\Lambda\to \proton\pim$ decays in the mass range $1.112\gev/c^2<M_{\proton\pim}<1.119\gev/c^2$ (corresponding to a ${\sim3\sigma}$ range around the nominal $m_\Lambda$~\cite{Zyla:2020zbs}) and selected using $\Lambda$-momentum dependent criteria based on four parameters: the distance between the two decay product tracks at their closest approach, in the direction opposite that of the $\ep$ beam; the minimum distance between the decay product tracks and the interaction point (IP) in the transverse plane; the angular difference between the $\Lambda$ flight direction and the direction pointing from the IP to the $\Lambda$ decay vertex in the transverse plane; and the flight length of the $\Lambda$ in the transverse plane. Measurements from CDC, TOF, and ACC are combined to form the charged particle identification (PID) likelihoods ${\cal L}(h)$ ($h=p, K\text{, or }\pi$), where ${\cal L}({h:h^\prime})$, defined as ${\cal L}(h)/\left[{\cal L}(h)+{\cal L}(h^\prime)\right]$, is the discriminator between the $h$ and $h^\prime$ hypotheses. We require ${\cal L}({p:K})>0.6$ and ${\cal L}({p:\pi})>0.6$ for the proton from the $\Lambda$ decay. The proton PID efficiency in the produced \BToLDM MC events is $70$--$99\%$, with the purity being $100\%$.

To suppress background with particles undetected along the beampipe, we require the cosine of the polar angle of the missing momentum in the laboratory frame~\cite{Belle} to lie between $-0.86$ and $0.95$. After identifying the $B_\text{tag}$ candidate and reconstructing the $\Lambda$ baryon, we require that no additional charged tracks remain in the event. This veto and the PID requirements for protons eliminate any multiple $\Lambda$ candidates in our samples. If there are multiple $B_\text{tag}$ candidates in an event, the candidate with the highest $o_\text{tag}$ value is chosen.

While the topological distribution of particles in \B decay events is more isotropic, that for continuum events, the dominant background at this level, is more jet-like. To suppress the continuum background, we simultaneously optimize the requirements on two event-shape variables: the ratio of the second to zeroth Fox--Wolfram moments~\cite{Fox:1978vu} of the event, $R_2$; and the cosine of the angle between the thrust axis~\cite{Brandt:1964sa} of the $\Lambda$ and the thrust axis of the $B_\text{tag}$, $\cos\theta_\text{T}$. This is done by maximizing a figure-of-merit $\varepsilon/(\frac{\alpha}{2}+\sqrt{N_B})$, which is independent of the assumed signal branching fraction and optimized for new decay modes~\cite{Punzi:2003bu}. Here, $\varepsilon$ is the signal efficiency while $N_B$ denotes the number of background events passing the requirements on the two event-shape variables. Both values are determined from MC simulation, with the number of continuum events being corrected with an overall factor based on the off-resonance data. We choose $\alpha=3$, the number of standard deviations of the desired sensitivity.

\begin{figure}[htb]
\includegraphics[width=0.45\textwidth]{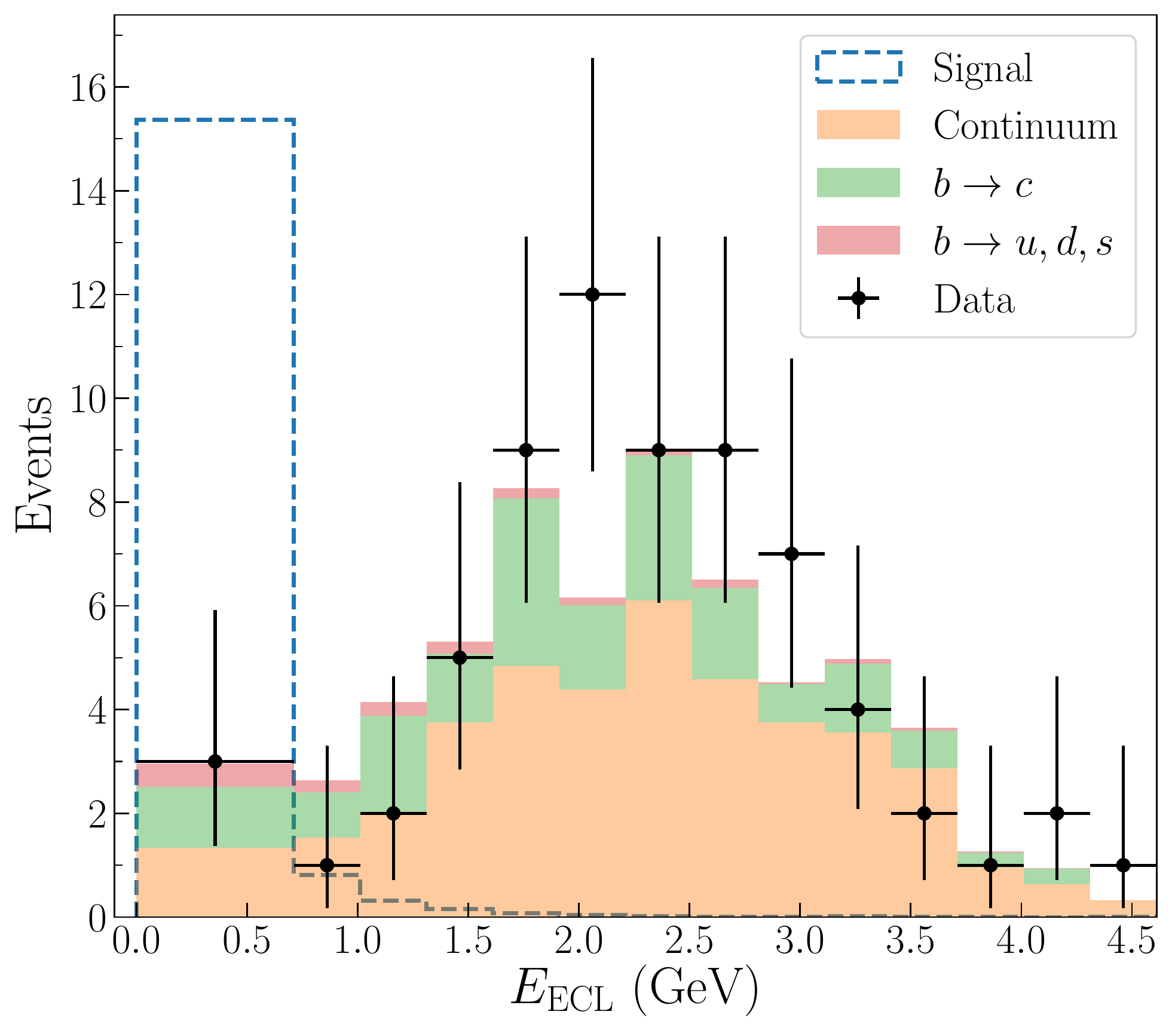}
\caption{The observed (solid points) and expected background (stacked shaded histograms) $E_{\text{ECL}}$ distributions for $m_{\psiDM}=2.5\gev/c^2$, with the first bin representing the signal region. The shape of the continuum contribution is taken from the off-resonance data, while the other two background sources are from MC simulation; each distribution is normalized to the expected number of events in the first bin. The signal shape (dashed line) is taken from MC simulation, assuming a branching fraction of $\BFDM=8\times 10^{-5}$.}
\label{fig:ecl}
\end{figure}

The most powerful variable to identify signal decays is the residual energy in the ECL, $E_{\text{ECL}}$, which is the sum of the energies of ECL clusters that are not associated with the $B_\text{tag}$ decay products nor with the signal-side $\Lambda$ candidate. To suppress contributions from noise in the ECL, minimum thresholds on the cluster energy are required: $50\mev$ for the barrel, $100\mev$ for the forward endcap, and $150\mev$ for the backward endcap region. These thresholds were determined to achieve an optimal signal-to-noise ratio in the calorimeter clusters. The decays $\B\to\Dstar l\nu$ are examined as control samples; the observed $E_{\text{ECL}}$ distributions are found to be in good agreement with MC simulations. In a correctly reconstructed signal event, no additional activity should appear in the calorimeter, so the $E_{\text{ECL}}$ distribution for signal events will peak at low $E_{\text{ECL}}$ values. The $E_{\text{ECL}}$ signal region is selected from MC simulation. It is defined by requiring that the expected number of background events corresponding to the full Belle dataset is $\cong 3$, thus allowing downward fluctuation of the number of observed events, $n_\text{obs}$. The contamination of continuum background events in the $E_{\text{ECL}}$ signal region is determined from MC simulation, with the scale factor being determined with the off-resonance data. In addition, a correction for possible background from rare \B decays is applied to the number of \BB events. After all the selection requirements are applied, approximately half of the expected background consists of continuum events and the other half from \BB events. The observed and expected background $E_{\text{ECL}}$ distributions for $m_{\psiDM}=2.5\gev/c^2$ are shown in Fig.~\ref{fig:ecl}, together with the corresponding signal distribution.

The signal yield is obtained by counting the number of events in the signal region. The profile likelihood method~\cite{Cowan:2010js} is used to calculate the $90\%$ confidence level (CL) upper limits on \BFDM, as a function of the $\psiDM$ mass. The likelihood is a Poisson ``on/off" model with log-normal constraints to incorporate systematic uncertainties. The signal yield is defined as $s=2\times N_{\BzBzb} \times \epsilon \times \BFDM \times \BFLs$, where $ N_{\BzBzb}$ is the number of $\BzBzb$ pairs in the full Belle dataset, $\epsilon$ is the signal efficiency, and $\BFLs$ denotes the branching fraction $\BFL$. The likelihood is defined as
\begin{linenomath}
\begin{align}
\label{eq:ln}
{\cal L} & = \mathcal{P}(n_\text{sr}; s + b)  \times \mathcal{P}(n_\text{bkg};\tau\cdot b_0) \nonumber \\ 
& \times \mathcal{G}\left[b;b_0,1+\textsl{syst}(b_0)\right] \nonumber \\ 
& \times \mathcal{G}\left[\epsilon;\textsl{nomi}(\epsilon),1+\textsl{syst}(\epsilon)\right] \nonumber \\ 
& \times \mathcal{G}\left[N_{\BzBzb};\textsl{nomi}(N_{\BzBzb}),1+\textsl{syst}(N_{\BzBzb})\right] \nonumber \\ 
& \times \mathcal{G}\left[\BFLs;\textsl{nomi}(\BFLs),1+\textsl{syst}(\BFLs)\right].
\end{align}
\end{linenomath}
Here, $\mathcal{P}$ is a Poisson distribution, and
\begin{linenomath}
\begin{equation}
\mathcal{G}(x;m_0,\kappa)=\frac{1}{x\sqrt{2\pi}\ln(\kappa)}\exp\left[{-\frac{\ln^2(x/m_0)}{2\ln^2(\kappa)}}\right]
\end{equation}
\end{linenomath}
is a log-normal distribution, where $m_0$ is the median identified with the best estimate for the random variable $x$, and $\kappa>1$ encodes the spread in the distribution with $\kappa-1$ corresponding roughly to the multiplicative relative uncertainty on $x$. In Eq.~(\ref{eq:ln}), $n_\text{sr}$ is the number of candidates in the signal region, $n_\text{bkg}$ is the expected number of background MC candidates surviving our selection criteria, $\tau=6$ is the ratio between the luminosity of the background MC sample and the full Belle dataset, $\textsl{syst}$ denotes the relative systematic uncertainty and $\textsl{nomi}$ the nominal value. The expected number of background events corresponding to the full Belle dataset, $b_0$, is a free parameter of the likelihood. It has the same relative systematic uncertainty as $n_\text{bkg}$.

For each $m_{\psiDM}$ value we allow different optimization requirements, and different systematic uncertainties are included based on the kinematics and reconstruction efficiency of each sample. The systematic uncertainty arising from the number of \BB pairs is $1.4\%$. The world average value of ${\cal B}\left[\FourS\rightarrow \BzBzb\right]$ is $(48.6\pm0.6)\%$~\cite{Zyla:2020zbs}, leading to a systematic uncertainty of $1.8\%$ on the number of \BzBzb pairs. The world average value of ${\cal B}_{\Lambda}$ is $(63.9\pm0.5)\%$~\cite{Zyla:2020zbs}, resulting in the systematic uncertainty of $0.8\%$.

The calibration factor for the hadronic tagging efficiency is studied in Ref.~\cite{Keck:2018lcd} and found to be $0.860\pm0.074$. The uncertainty in this value is taken as a systematic uncertainty. The uncertainty due to proton PID is evaluated using an independent sample of $\Lambda\to \proton\pim$ decays. The systematic uncertainty due to charged track reconstruction is calculated using partially reconstructed $\Dstarp \to\Dz\pip$ decays, with $\Dz\to\KS \pipi$ and $\KS\to\pipi$. For the pion, an additional systematic uncertainty based on a study of low-momentum tracks from $\Bz\to\Dstarm\pip$ decays is applied. The slow pion emitted in the decays of the $\Dstar$ allows one to probe the low-momentum region. The difference in the charged track veto efficiency between data and MC simulation is estimated by comparing the effect of requiring no extra tracks available in the event on samples of $\Bz\to\ensuremath{D^{(*)}}\xspace l\nu$ events. The systematic uncertainty due to $\Lambda$ reconstruction is determined from a comparison of yield ratios of $\Bp\to\Lambda\Lbar\Kp$ with and without the $\Lambda$ selection requirements in data and MC samples. The weighted average of the data-MC difference over the momentum range is assigned as the systematic uncertainty. Furthermore, we include the binomial error of the efficiency as a systematic uncertainty.

The statistical uncertainty of the correction on the number of continuum events in MC simulation is $21.0$--$26.9\%$, based on off-resonance data; this is assigned as a systematic uncertainty. Since none of the \BB background MC decays surviving our selection criteria are from exclusively observed and measured processes in experiment, we assign a conservative $50\%$ systematic uncertainty on their branching fractions; this is the dominant systematic uncertainty on $n_\text{bkg}$. The statistical uncertainty of the correction for possible background from rare \B decays applied to the number of \BB background events in MC simulation is included as a systematic uncertainty.

\begin{figure}[htb]
\includegraphics[width=0.45\textwidth]{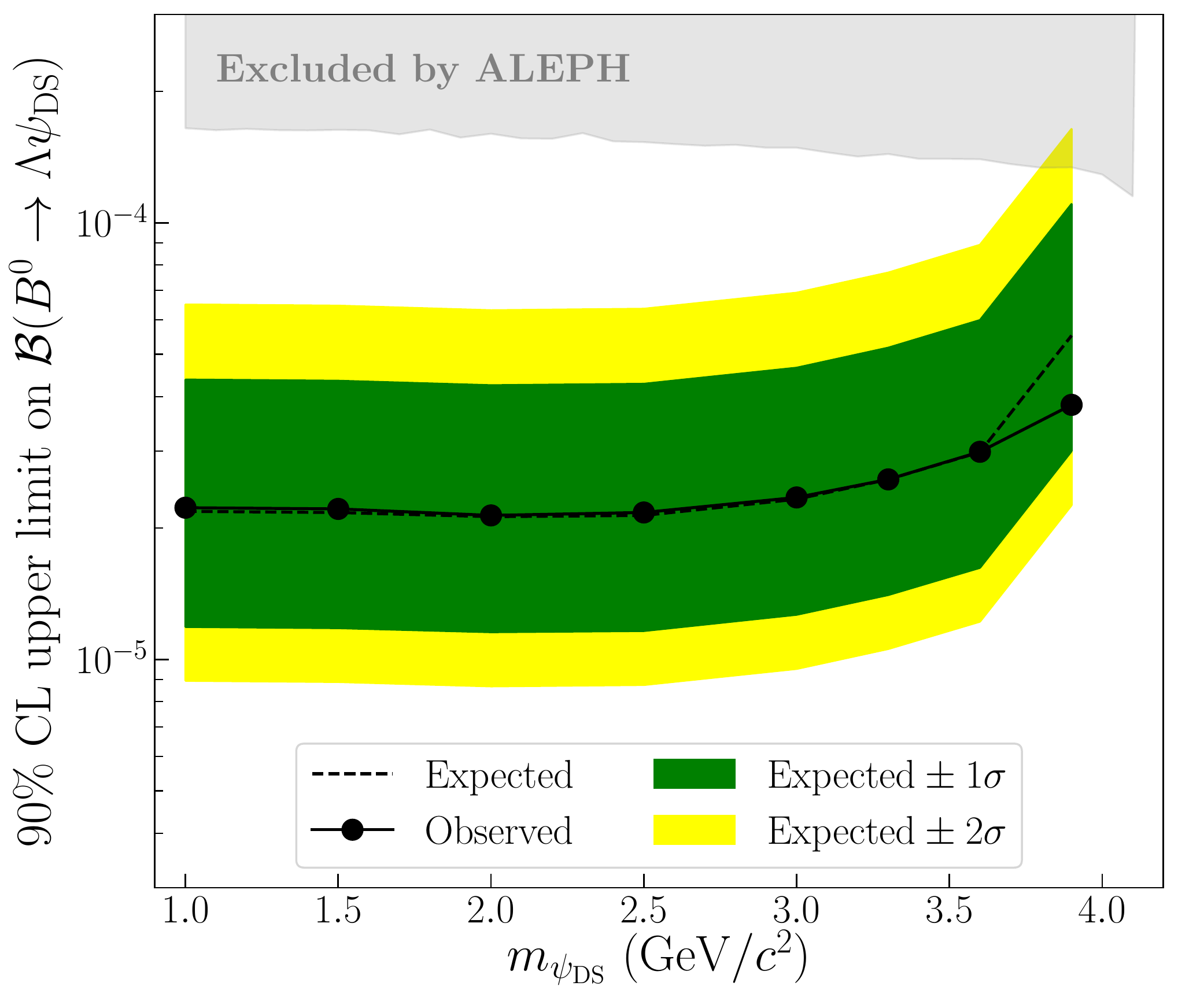}
\caption{The observed (solid line) and median expected (dashed line) $90\%$ CL upper limits on \BFDM as a function of $m_{\psiDM}$. The $\pm 1\sigma$ and $\pm 2\sigma$ expected exclusion regions are indicated in green and yellow, respectively. A linear interpolation is performed between the values obtained for the probed $m_{\psiDM}$ values. The gray shaded region shows the resulting $90\%$ CL constraints from the reinterpretation of a search at ALEPH for decays of $b$-flavored hadrons with large missing energy~\cite{ALEPH, Alonso-Alvarez:2021qfd}.}
\label{fig:obsul}
\end{figure}

The range of systematic uncertainties in the estimate of the signal efficiencies, $\delta\epsilon$, and the number of expected $\BB$ background events, $\delta n_\text{bkg}^{\BB}$, across the different values of $m_{\psiDM}$ are listed in Table~\ref{tab:sys}. The observed and expected $90\%$ CL upper limits on \BFDM as a function of $m_{\psiDM}$ are shown in Fig.~\ref{fig:obsul}. A summary of these limits and the different distinct variables used in their calculation for each $m_{\psiDM}$ is presented in Table~\ref{tab:summary}.

\begin{table}[htb]
\caption{Range of systematic uncertainties in the estimate of the signal efficiencies, $\delta\epsilon$, and the number of expected $\BB$ background events, $\delta n_\text{bkg}^{\BB}$, across the different values of $m_{\psiDM}$.}
\label{tab:sys}
\begin{tabular}
 {@{\hspace{0.4cm}}l@{\hspace{0.4cm}}  @{\hspace{0.4cm}}c@{\hspace{0.4cm}} @{\hspace{0.4cm}}c@{\hspace{0.4cm}}}
\hline \hline
Source &  $\delta\epsilon$ (\%) & $\delta n_\text{bkg}^{\BB}$ (\%) \\
\hline
$B_\text{tag}$ correction & 8.6 & 8.6\\
Proton PID & 0.5--2.8 & 4.3--5.7 \\
Tracking efficiency  & 0.7--1.9 & 1.1--1.9 \\
Charged track veto & 5.3--6.5 & 5.3--6.5 \\
$\Lambda$ selection & 2.5--3.6 & 4.4--4.7 \\
Signal MC statistics & 1.2--2.0 & --\\
Rare \B decays correction & -- & 10.6--13.4 \\
Branching fractions & -- & 50.0 \\
\hline \hline
\end{tabular}
\end{table}

\begin{table*}[htb]
\caption{Summary of the requirements on the event-shape variables $R_2$ and $\cos\theta_\text{T}$, the $E_\text{ECL}$ signal region, the signal efficiency $\epsilon$ and its systematic uncertainty $\delta\epsilon$, the systematic uncertainty on the expected number of background MC candidates surviving our selection criteria $\delta n_\text{bkg}$, the expected $90\%$ CL upper limit, the number of observed events $n_\text{obs}$, and the observed $90\%$ CL upper limit, for each $m_{\psiDM}$.}
\label{tab:summary}
\begin{tabular}
{@{\hspace{0.3cm}}c@{\hspace{0.3cm}}  @{\hspace{0.3cm}}c@{\hspace{0.3cm}} @{\hspace{0.3cm}}c@{\hspace{0.3cm}} @{\hspace{0.3cm}}c@{\hspace{0.3cm}} @{\hspace{0.3cm}}c@{\hspace{0.3cm}} @{\hspace{0.3cm}}c@{\hspace{0.3cm}}  @{\hspace{0.3cm}}c@{\hspace{0.3cm}} @{\hspace{0.3cm}}c@{\hspace{0.3cm}} @{\hspace{0.3cm}}c@{\hspace{0.3cm}} @{\hspace{0.3cm}}c@{\hspace{0.3cm}}}
\hline \hline
\thead{$m_{\psiDM}$\\(${\ensuremath{\mathrm{Ge\kern -0.1em V}}\xspace}/c^2$)} & $R_2$ & $\cos\theta_\text{T}$ & \thead{$E_\text{ECL}$\\(${\ensuremath{\mathrm{Ge\kern -0.1em V}}\xspace}$)} &$\epsilon$ ($10^{-4}$) & $\delta\epsilon$ ($\%$) & $\delta n_\text{bkg}$ ($\%$) & \thead{Expected\\limit ($10^{-5}$)} & $n_\text{obs}$ & \thead{Observed\\limit ($10^{-5}$)} \\
\hline
1.0 & $<0.31$ & $<0.66$ & $<0.74$ & 3.92 & 11.0 & 30.3 & 2.2 & 3 & 2.2 \\
1.5 & $<0.30$ & $<0.66$ & $<0.74$ & 3.94 & 11.0 & 30.3 & 2.2 & 3 & 2.2 \\
2.0 & $<0.31$ & $<0.70$ & $<0.74$ & 4.05 & 10.9 & 30.6 & 2.1 & 3 & 2.1 \\
2.5 & $<0.33$ & $<0.67$ & $<0.71$ & 4.01 & 10.9 & 30.7 & 2.1 & 3 & 2.2 \\
3.0 & $<0.33$ & $<0.70$ & $<0.71$ & 3.69 & 11.0 & 30.8 & 2.3 & 3 & 2.3 \\
3.3 & $<0.35$ & $<0.70$ & $<0.68$ & 3.32 & 11.1 & 28.4 & 2.6 & 3 & 2.6 \\
3.6 & $<0.44$ & $<0.70$ & $<0.63$ & 2.88 & 11.7 & 27.7 & 3.0 & 3 & 3.0 \\
3.9 & $<0.42$ & $<0.79$ & $<0.57$ & 1.56 & 11.3 & 30.2 & 5.5 & 2 & 3.8 \\
\hline \hline
\end{tabular}
\end{table*}

\begin{figure}[htb]
\includegraphics[width=0.45\textwidth]{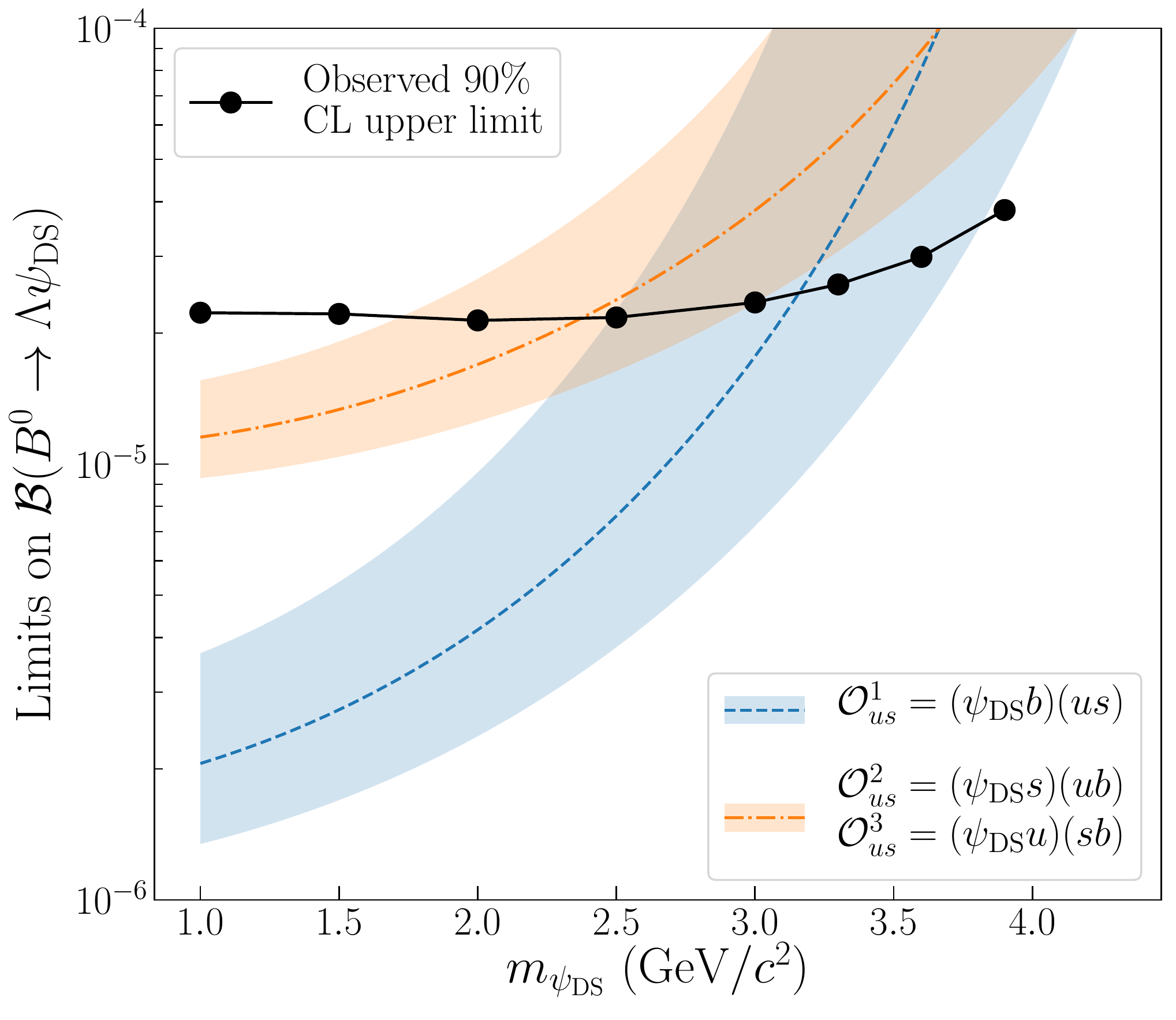}
\caption{The observed $90\%$ CL upper limits on \BFDM as a function of $m_{\psiDM}$ (solid line), and the lower bounds on \BFDM for $B$-Mesogenesis using phase-space considerations (shaded bands). The $b$-quark pole mass is chosen as the benchmark mass in the phase-space integral (dashed lines) while two other choices, the \Bz meson mass and the $b$-quark $\overline{\text{MS}}$ mass, delineate the upper and lower edges of the shaded bands, respectively. The calculation is performed for the ``type-1" operator ${\cal O}_{us}^1=(\psiDM b)(us)$, and the ``type-2" and ``type-3" cases ${\cal O}_{us}^2=(\psiDM s)(ub)$ and ${\cal O}_{us}^3=(\psiDM u)(sb)$, for which the phase-space integration is the same.}
\label{fig:obsulth}
\end{figure}

The fraction of decays not expected to contain hadrons other than $\Lambda$ in the final state as a function of $m_{\psiDM}$ is calculated in Ref.~\cite{Alonso-Alvarez:2021qfd} using phase-space considerations. This fraction multiplied with ${\cal B}_\text{M}$ provides the lower bounds on \BFDM for $B$-Mesogenesis. Those bounds together with the observed $90\%$ CL upper limits on \BFDM as a function of $m_{\psiDM}$ are presented in Fig.~\ref{fig:obsulth}. The region $m_{\psiDM}\gtrsim3.0\gev/c^2$ is excluded for the ${\cal O}_{us}^2$ and ${\cal O}_{us}^3$ operator cases.

In summary, we have reported the results of a search for the decays of \Bz mesons into a final state containing a $\Lambda$ baryon and missing energy with a fully reconstructed $B_\text{tag}$ using a data sample of $772\times10^6$ \BB pairs collected at the \FourS resonance with the Belle detector. No significant signal is observed and we set upper limits on the branching fractions at $90\%$ CL, which are the most stringent constraints to date. Our analysis yields significant improvements, and partially excludes the $B$-Mesogenesis mechanism. We expect that the Belle II experiment~\cite{Abe:2010gxa} will be able to fully test this mechanism.

The authors would like to thank G.~Alonso-\'{A}lvarez, G.~Elor, M.~Escudero, and A.~Nelson for useful discussions on the \B-Mesogenesis mechanism.
We thank the KEKB group for the excellent operation of the
accelerator; the KEK cryogenics group for the efficient
operation of the solenoid; and the KEK computer group, and the Pacific Northwest National
Laboratory (PNNL) Environmental Molecular Sciences Laboratory (EMSL)
computing group for strong computing support; and the National
Institute of Informatics, and Science Information NETwork 5 (SINET5) for
valuable network support.  We acknowledge support from
the Ministry of Education, Culture, Sports, Science, and
Technology (MEXT) of Japan, the Japan Society for the 
Promotion of Science (JSPS), and the Tau-Lepton Physics 
Research Center of Nagoya University; 
the Australian Research Council including grants
DP180102629, 
DP170102389, 
DP170102204, 
DP150103061, 
FT130100303; 
Austrian Federal Ministry of Education, Science and Research (FWF) and
FWF Austrian Science Fund No.~P~31361-N36;
the National Natural Science Foundation of China under Contracts
No.~11435013,  
No.~11475187,  
No.~11521505,  
No.~11575017,  
No.~11675166,  
No.~11705209;  
Key Research Program of Frontier Sciences, Chinese Academy of Sciences (CAS), Grant No.~QYZDJ-SSW-SLH011; 
the  CAS Center for Excellence in Particle Physics (CCEPP); 
the Shanghai Pujiang Program under Grant No.~18PJ1401000;  
the Shanghai Science and Technology Committee (STCSM) under Grant No.~19ZR1403000; 
the Ministry of Education, Youth and Sports of the Czech
Republic under Contract No.~LTT17020;
Horizon 2020 ERC Advanced Grant No.~884719 and ERC Starting Grant No.~947006 ``InterLeptons'' (European Union);
the Carl Zeiss Foundation, the Deutsche Forschungsgemeinschaft, the
Excellence Cluster Universe, and the VolkswagenStiftung;
the Department of Atomic Energy (Project Identification No. RTI 4002) and the Department of Science and Technology of India; 
the Istituto Nazionale di Fisica Nucleare of Italy; 
National Research Foundation (NRF) of Korea Grant
Nos.~2016R1\-D1A1B\-01010135, 2016R1\-D1A1B\-02012900, 2018R1\-A2B\-3003643,
2018R1\-A6A1A\-06024970, 2018R1\-D1A1B\-07047294, 2019K1\-A3A7A\-09033840,
2019R1\-I1A3A\-01058933;
Radiation Science Research Institute, Foreign Large-size Research Facility Application Supporting project, the Global Science Experimental Data Hub Center of the Korea Institute of Science and Technology Information and KREONET/GLORIAD;
the Polish Ministry of Science and Higher Education and 
the National Science Center;
the Ministry of Science and Higher Education of the Russian Federation, Agreement 14.W03.31.0026, 
and the HSE University Basic Research Program, Moscow; 
University of Tabuk research grants
S-1440-0321, S-0256-1438, and S-0280-1439 (Saudi Arabia);
the Slovenian Research Agency Grant Nos. J1-9124 and P1-0135;
Ikerbasque, Basque Foundation for Science, Spain;
the Swiss National Science Foundation; 
the Ministry of Education and the Ministry of Science and Technology of Taiwan;
and the United States Department of Energy and the National Science Foundation.

\end{document}